\documentclass[aps,prb,showpacs,twocolumn,amsmath,amssymb,superscriptaddress,letterpaper]{revtex4}
\usepackage{times}
\usepackage{amsfonts}
\usepackage{mathrsfs}
\usepackage{graphicx}% Include figure files
\usepackage{dcolumn}% Align table columns on decimal point
\usepackage{bm}% bold math
\usepackage{color}

\usepackage[colorlinks,bookmarks=false,citecolor=blue,linkcolor=red,urlcolor=blue]{hyperref}
\def\be{\begin{equation}}       \def\ee{\end{equation}}
\def\bea{\begin{eqnarray}}      \def\eea{\end{eqnarray}}

\begin{document}
\title{Three-dimensional Topological Critical Dirac semimetal in $A$MgBi ($A$=K, Rb, Cs)}

\author{Congcong Le}\thanks{these authors contributed equally to this paper}
\affiliation{ Institute of Physics, Chinese Academy of Sciences,
Beijing 100190, China}

\author{Shengshan Qin}\thanks{these authors contributed equally to this paper}
\affiliation{ Institute of Physics, Chinese Academy of Sciences,
Beijing 100190, China}

\author{Xianxin Wu}
\affiliation{ Institute of Physics, Chinese Academy of Sciences,
Beijing 100190, China}\affiliation{Institute for Theoretical Physics and Astrophysics,
Julius-Maximilians University of W¡§urzburg, Am Hubland, D-97074 W¡§urzburg, Germany}

\author{Xia Dai}
\affiliation{ Institute of Physics, Chinese Academy of Sciences,
Beijing 100190, China}

\author{Peiyuan Fu}
\affiliation{ Institute of Physics, Chinese Academy of Sciences,
Beijing 100190, China}

\author{Chen Fang}
\affiliation{ Institute of Physics, Chinese Academy of Sciences,
Beijing 100190, China}

\author{Jiangping Hu  }\email{jphu@iphy.ac.cn} \affiliation{
Institute of Physics, Chinese Academy of Sciences, Beijing 100190,
China}\affiliation{Collaborative Innovation Center of Quantum Matter, Beijing, China}

\date{\today}

\begin{abstract}

We predicted, using first principles calculation, that $A$MgBi ($A$=K, Rb, Cs) are symmetry-protected topological semimetals near the boundary of type-\uppercase\expandafter{\romannumeral1} and type-\uppercase\expandafter{\romannumeral2} Dirac semimetal phases, dubbed topological critical Dirac semimetals. Doping Rb or Cs into KMgBi can drive the transition between the two phases. An effective theory is developed to describe the bands near the Fermi energy, by which we calculate the surface Fermi arcs and the Landau levels throughout the transition. We predict the key features of critical Dirac semimetals that can be observed in photoemission, quantum oscillation and transport measurements.

\end{abstract}


\maketitle

Topological semimetals (TSs) are semimetals whose Fermi surfaces carry nontrivial topological numbers. These quantum numbers lead to a series of exotic effects such as the existence of Fermi arcs on the surface \cite{HgCrSe,XG Wang} and the chiral anomaly \cite{Burkov,CAE} in the bulk transport. Crystal symmetries afford us a large variety of topological semimetals including Weyl semimetals (WS) \cite{intermediate phase,multilayer weyl,HgCrSe,xu2015,Lv2015,Weng2015,Shekhar2015,Yang2015,Xu2015}, Dirac semimetals (DS) \cite{Dirac Kane,Na3Bi,Cr3As2} and nodal line semimetals \cite{line node1,line node2}. For Weyl/Dirac semimetals, a further distinction has been made between the type-I and the type-II classes \cite{type2}, where the Fermi surfaces (at ideal half-filling) are point-like and pocket-like, respectively. The physical consequences of both types have been studied \cite{magneto1,magneto2}.

For its unique feature and potential application, TSs have drawn great attention in the field of topological materials. In recent years, great progress has been made both theoretically and experimentally. Na$_3$Bi and Cd$_3$As$_2$ have been predicted to be three-dimensional (3D) linear DSs theoretically \cite{Na3Bi,Cr3As2} and  have been verified by angle-resolved photoemission spectroscopy (ARPES) measurements \cite{Na3Bi ARPES,Cr3As2 ARPES}. In transport experiments, Cd$_3$As$_2$ and several other DSs exhibit  strong linear magnetoresistance\cite{Cr3As2 MR,BiSb MR},  which is also a strong evidence for DSs.

In this Letter, we propose, by using first principles calculation, a family of materials, $A$MgBi, as topological semimetals that lie in between the type-I and the type-II Dirac semimetals, where $A$ is an alkaline metal ($A$=K, Rb, Cs). While KMgBi, which has been synthesized\cite{Rainer1979}, is a type-I Dirac semimetal, both RbMgBi and CsMgBi, are both type-II Dirac semimetals. Doping Rb and Cs into KMgBi hence drives a transition from type-I to type-II Dirac semimetals. Hence we call these compounds topological critical Dirac semimetals, which can help us understand this topological phase transition.

To further study the physical observables in these compounds, we develop an effective model that captures the key features near the Fermi energy throughout the topological transition. Using the model, we computed the surface states with Fermi arcs and also the Landau levels in the presence of magnetic field. These results can be directly observed in ARPES, Scanning Tunneling Microscope (STM) and transport measurements.

{\it Method and Crystal Structure}
Our calculations are performed using density functional theory (DFT) as implemented in the Vienna ab initio simulation package (VASP) code \cite{Kresse1993,Kresse1996,Kresse1996B}. The Perdew-Burke-Ernzerhof (PBE) exchange-correlation functional and the projector-augmented-wave (PAW) approach are used. Throughout the work, the cutoff energy is set to be 500 eV for expanding the wave functions into plane-wave basis. In the calculation, the BZ is sampled in the k space within Monkhorst-Pack scheme\cite{MonkhorstPack}. On the basis of the equilibrium structure, the k mesh used is $10\times10\times6$. We relax the lattice constants and internal atomic positions with GGA, where the plane wave cutoff energy is 600 eV. Forces are minimized to less than 0.01 eV/\AA in the relaxation.

The crystal structure of KMgBi \cite{Rainer1979}with the space group $P4/nmm$ is similar to the 111 family of iron-based superconductors\cite{Michael2008,Joshua2008,Dinah2009}. The Mg$_2$Bi$_2$ layers possess an anti-PbO-type atom arrangement, consisting of a square lattice sheet of Mg coordinated by Bi above and below the plane to form face sharing MgBi$_4$ tetrahedra. Details about the optimized and experimental structural parameters are summarized in the supplementary materials. We find that the optimized structural parameters are within 2\% from the corresponding experimental data. The bond angle of MgBi$_4$ tetrahedra is very close to that of the perfect tetrahedron. To further test the stability of optimized structural parameters, we calculate dispersions by using PHONOPY code\cite{Togo2008,Togo}(see Supplementary materials). No imaginary frequencies are observed throughout the whole Brillouin zone in phonon dispersions, confirming its dynamically structural stability. Therefore, we adopt the optimized structural parameters in the following calculations.

\begin{figure}
\centerline{\includegraphics[width=0.5\textwidth]{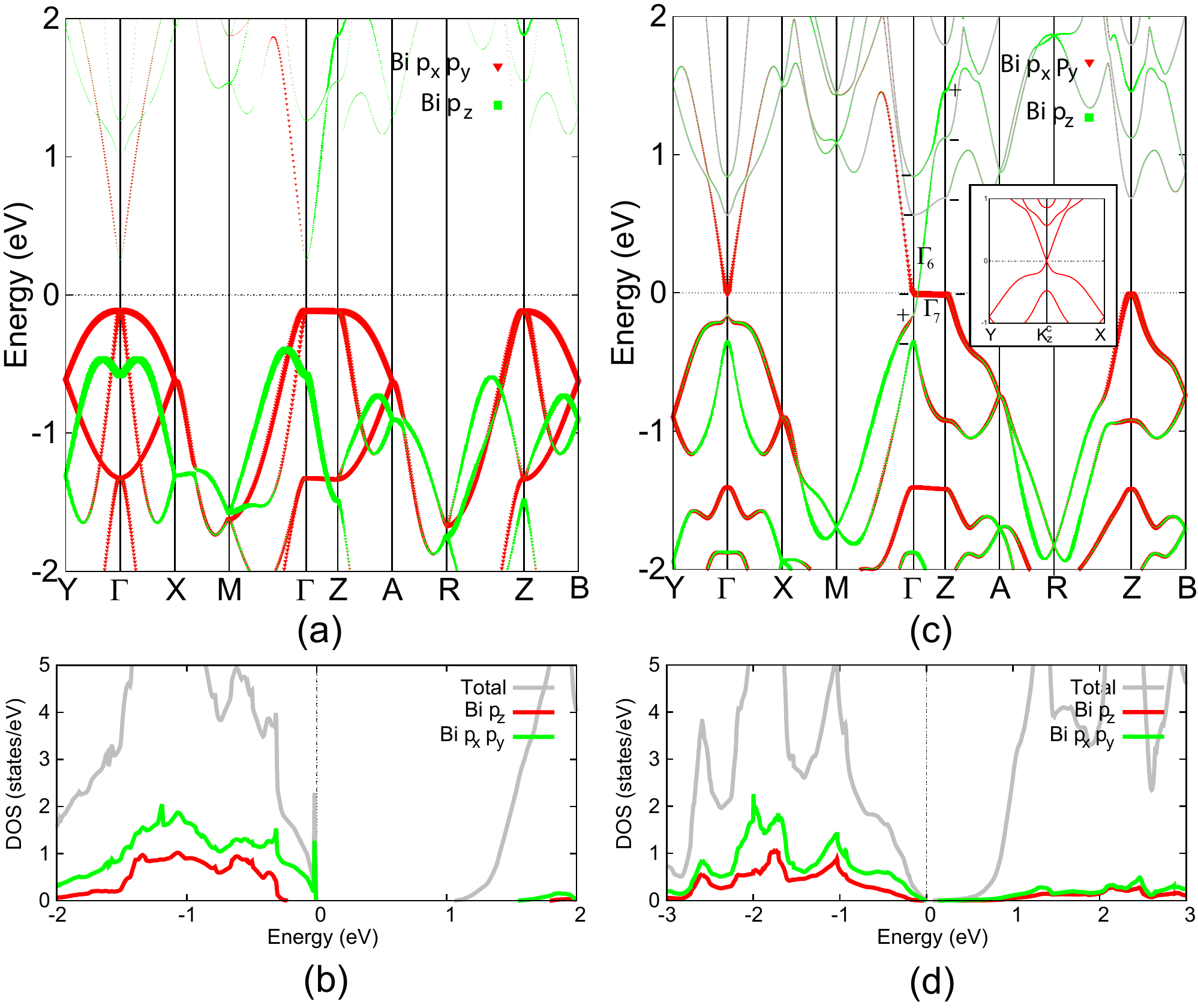}}
\caption{(color online) (a) and (b) are band structure and DOS of KMgBi without SOC. (c) and (d) are band structure and DOS of KMgBi with SOC.  The orbital characters of bands are represented by different colors. The parities of the eigenstates and the irreducible representations they belong to at $\Gamma$ point near the Fermi level are represented by $\pm$ and $\Gamma_i$ respectively. The inset is Band structure of KMgBi in the $k_z=k_z^{c}$ plane.
\label{banddos} }
\end{figure}

{\it Electronic structure}
The band structure and density of states (DOS) of KMgBi with GGA optimized structural parameters are displayed in Fig.\ref{banddos}. The paramagnetic state is insulating with an energy gap of 362.8 meV, as shown in Fig.\ref{banddos}(a). As spin orbital coupling (SOC) is large in Bi atoms, when SOC is included, the band inversion occurs around $\Gamma$ point as shown in Fig.\ref{banddos}(c). Near the Fermi level, the valence and conduction bands are mainly attributed to the Bi-6p orbitals. The band inversion happens between Bi-6p$_{x,y}$ and Bi-6p$_z$ orbitals, which is different from the known DS Na$_3$Bi\cite{Na3Bi} where the inverted band structure is formed by the Na-3s and Bi-6p$_{x,y}$ orbitals near the Fermi level. To further confirm the topological property in KMgBi, due to the presence of inversion symmetry, we can  adopt the parity check method proposed by Fu and Kane\cite{Fu2007} to calculate the topological invariance. The parities of the eigenstates at $\Gamma$ and Z points near the Fermi level are displayed in Fig.\ref{banddos}(c). The odd parity states are attributed to the Bi-6p$_{x,y}$ orbitals, while the even parity states are contributed by Bi-6p$_z$ orbital. Therefore, the band structure is topologically nontrivial according to the calculation of parity product of occupied states at the time reversal invariant momenta.

 When SOC is turned on, the inverted bands of KMgBi near the Fermi level do not reopen a gap. Thus, it is a Dirac semimetal with two Dirac nodes located at (0, 0, $\pm$0.1345$\times\frac{\pi}{c}$) along the $\Gamma$-Z line exactly at the Fermi level, and the Fermi surfaces of KMgBi are composed of two isolated Fermi points. To further confirm the band crossings, the irreducible representations of the bands along the $\Gamma$-Z line are calculated, shown in Fig.\ref{banddos}(c). The two crossing bands belong to $\Gamma_6$ and $\Gamma_7$ irreducible representations respectively, indicating that the crossing points cannot open a gap. Furthermore, due to the presence of both time reversal and inversion symmetries  in KMgBi,  each of the crossing points  has a four-fold degeneracy and is robustly protected by these symmetries.

\begin{figure}
\centerline{\includegraphics[width=0.5\textwidth]{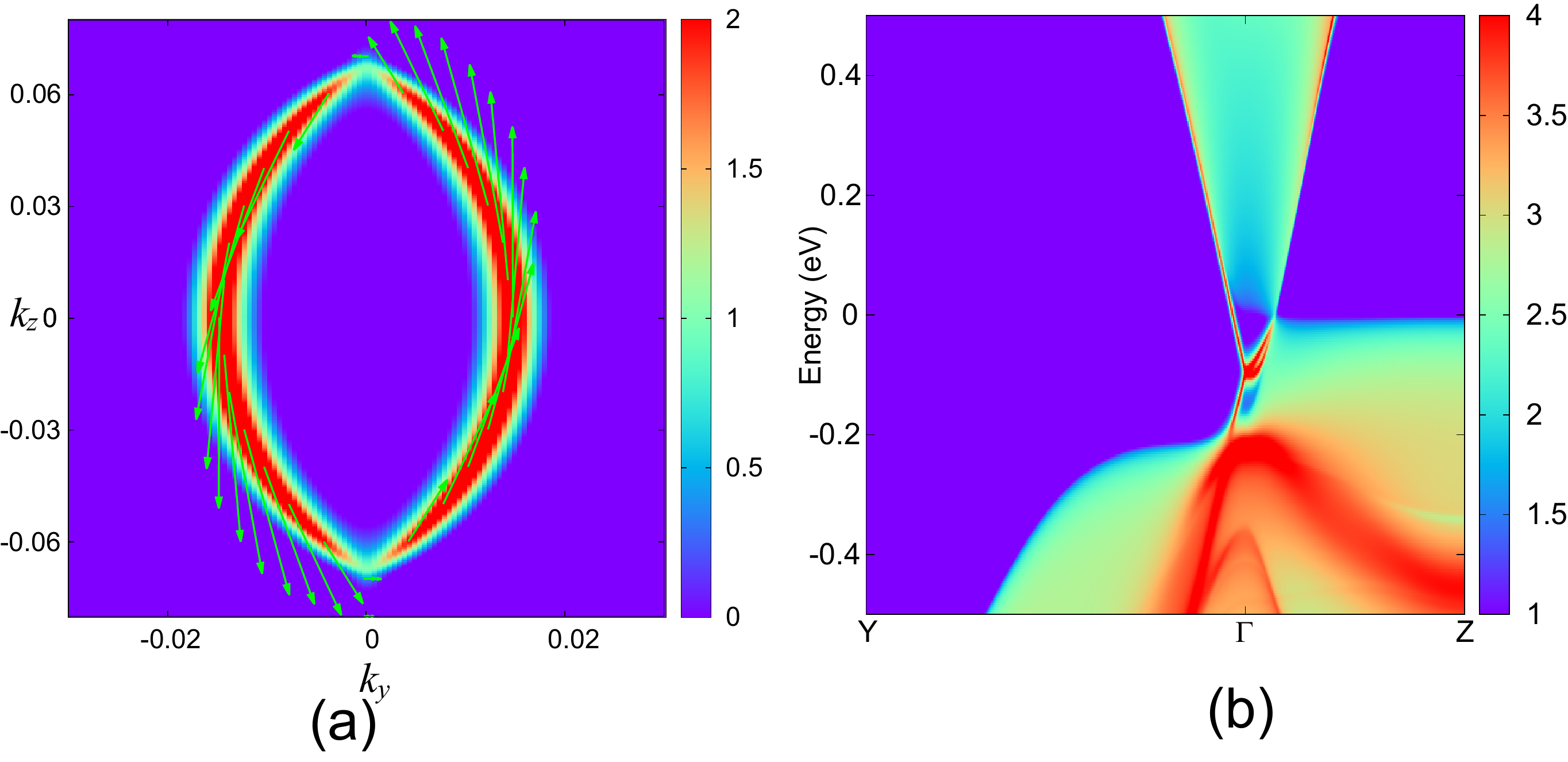}}
\caption{(color online) (a) The spin texture for the [100] surface states. (b) The projected surface states of KMgBi for [100] surface.
\label{edge} }
\end{figure}

The surface states of KMgBi can be obtained by calculating the surface Green function of the semi-infinite system using an iterative method\cite{Sancho1984,Sancho1985}. Fig.\ref{edge}(b) shows the edge states on the [100] surface, and the corresponding Fermi surface and in-plane spin texture are shown in Fig.\ref{edge}(a). Similar to the case of Na$_3$Bi\cite{Na3Bi}, on the [100] surface, the surface states are topologically nontrivial in the area $-k_z^c<k_z<k_z^c$ while it is topologically trivial in the area $|k_z|>k_z^c$. The corresponding Fermi surface on the [100] edge is a closed Fermi pocket composed of two Fermi arcs, and the two Fermi arcs connect each other at the projection points of the two Dirac points. Due to the spin-momentum lock, the spin texture on the the Fermi pocket is helical and it vanishes at the connecting points.

Now we address the lattice parameter effect on KMgBi. In principle, the lattice parameters can be changed by  external  pressure or by internal pressure created by chemical substitution. We have summarized the influence of the structural parameters on the gap size between the conduction band bottom and valence band top at $\Gamma$ point in the presence (absence) of SOC while the point group is kept unchanged in the supplementary materials. We find that, without SOC the gap at the $\Gamma$ point decreases with the increasing of structural parameters. This can be easily understood because   the energy gap between the bonding and anti-bonding states of Bi-6p$_{x,y,z}$ orbitals decreases with the increasing of lattice constants. In experiments, the lattice constants can also be tuned by doping Rb or Cs into the KMgBi, namely K$_{1-x}$R$_x$MgBi(R=Rb, Cs). In the supplementary materials, we have listed the optimized structural parameters and the energy gap between conduction band bottom and valence band top at $\Gamma$ point for KMgBi, K$_{0.5}$Rb$_{0.5}$MgBi and RbMgBi. It shows that with the increasing of Rb concentration both the lattice constants and the inverted gap increase.

\begin{figure}
\centerline{\includegraphics[width=0.5\textwidth]{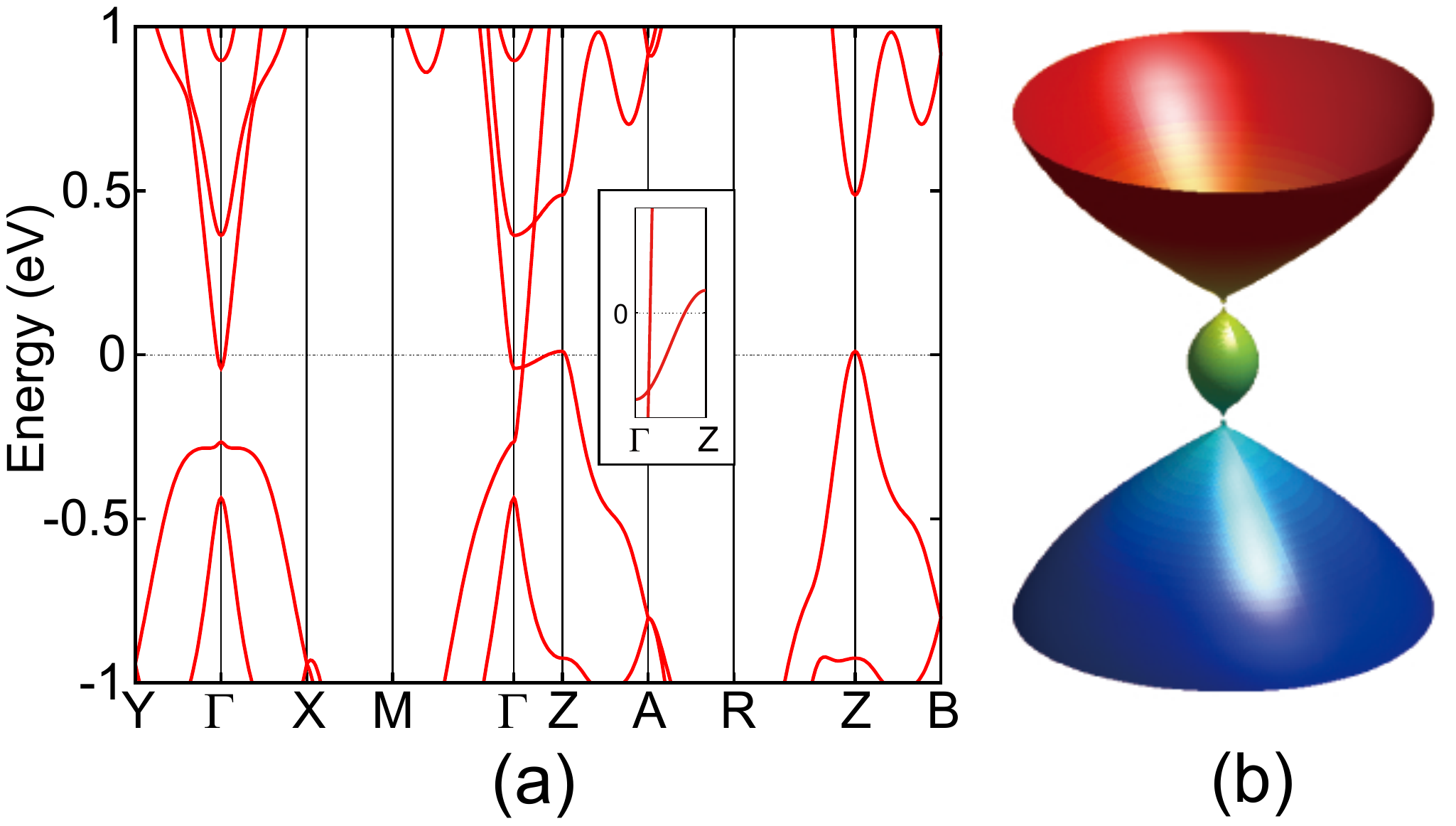}}
\caption{(color online) (a) Band structure of RbMgBi under the optimized structural parameters. Inset shows band structure of $\Gamma$-Z line. (b) Fermi surface of RbMgBi with fermi level tuned to -0.0375 eV. The color represents the $k_z$ value.
\label{band_fs} }
\end{figure}

In addition, for KMgBi, the dispersion of the $\Gamma_7$ band is very weak along the $\Gamma$-Z line. The energy of the $\Gamma_7$ band at $\Gamma$ point is only about $3.8$ meV higher than  the one at Z point, as shown in Fig.\ref{banddos}(c). Thus, KMgBi is located at the edge of type-\uppercase\expandafter{\romannumeral1} and type-\uppercase\expandafter{\romannumeral2} DS phases and it is a intriguing type-\uppercase\expandafter{\romannumeral1} DS. A small perturbation may drive KMgBi to be a type-\uppercase\expandafter{\romannumeral2} DS. To study the transition between type-\uppercase\expandafter{\romannumeral1} DS phase and type-\uppercase\expandafter{\romannumeral2} DS phase of KMgBi, we replace K with Rb atom. The band structures of RbMgBi shown in Fig.\ref{band_fs}(a) are very similar to those of KMgBi, except that the $\Gamma_7$ band of RbMgBi has a different dispersion along the $\Gamma$-Z line.  The $\Gamma_7$ band for RbMgBi at $\Gamma$ point is about 52 meV lower than the one at Z point and the inset in Fig.\ref{band_fs} shows band structure of $\Gamma$-Z line. The highly tilted conical dispersion around Dirac point clearly indicates that RbMgBi is a type-\uppercase\expandafter{\romannumeral2} DS.

{\it Effective Hamiltonian}
In order to get the effective Hamiltonian near the $\Gamma$ point where the band inversion occurs, we use the theory of invariants method in a similar way as for the Bi$_2$Se$_3$ family of materials\cite{Zhang2009,Liu2010}. From the band structure, the states around the $\Gamma$ point are mainly attributed to Bi-$p_{z}$ and Bi-$p_{x,y}$ orbitals, so we use these orbitals to construct the basis. Since Bi and Bi$^{\prime}$ atoms are related by the inversion symmetry, it is convenient to combine these orbitals to form the eigenstates of the inversion symmetry, which are given by

\begin{eqnarray}
 |P^{\mp}_{\alpha}\rangle=\frac{1}{\sqrt{2}}(|Bi_{\alpha}\rangle\pm|Bi^{\prime}_{\alpha}\rangle)
\end{eqnarray}
where the superscript denotes the parity and $\alpha=p_{x,y,z}$.

Taking SOC into account in the atomic picture, we can write the atomic Bi-6p states with SOC   as $|P^{\pm}_{\frac{3}{2}}, \pm\frac{3}{2}\rangle$,
$|P^{\pm}_{\frac{3}{2}}, \pm\frac{1}{2}\rangle$, $|P^{\pm}_{\frac{1}{2}}, \pm\frac{1}{2}\rangle$, where the subscript indicates the total angular momentum $J$. Due to the crystal field effect, $|P^{\pm}_{\frac{3}{2}},  \pm\frac{1}{2}\rangle$ and $|P^{\pm}_{\frac{1}{2}}, \pm\frac{1}{2}\rangle$ are  mixed  to form new eigenstates $|P^{\pm},\pm\frac{1}{2}\rangle$ that  define the energy spectra of the low energy model. Taking $|P^{+}, \frac{1}{2}\rangle$, $|P^{-}, \frac{3}{2}\rangle$, $|P^{+} , -\frac{1}{2}\rangle$, $|P^{-}, -\frac{3}{2}\rangle$ as the basis in $k\cdot{p}$ theory, we can  construct the effective Hamiltonian around the $\Gamma$ point as
\begin{equation}H_{0}(\mathbf{k})  = \epsilon_0(\mathbf{k})+ \left(\begin{array}{cccc}
M(\mathbf{k}) & ak_{-} & 0 & 0 \\
ak_{+} & -M(\mathbf{k}) & 0 & 0 \\
0 & 0 & M(\mathbf{k}) & -ak_{+} \\
0& 0 & -ak_{-} & -M(\mathbf{k})  \\
\end{array}\right),
\label{eq2}
\end{equation}
where $\epsilon_0(\mathbf{k})=C_0+C_1k^2_{z}+C_{2}(k^2_{x}+k^2_{y})$, $k_{\pm}=k_x{\pm}ik_y$ and
$M(\mathbf{k})=M_0-M_{1}k^2_{z}-M_{2}(k^2_{x}+k^2_{y})$. $M_0M_2<0$ corresponds to the topologically trivial regime while $M_0M_2>0$ corresponds to the band inversion regime, namely, the topologically nontrivial regime. The energy spectrum of the $k\cdot{p}$ Hamiltonian is
\begin{eqnarray}
E(\mathbf{k})=\epsilon_0(\mathbf{k})\pm\sqrt{M(\mathbf{k})^2+a^2k_{+}k_{-}},
\end{eqnarray}
where each band here is two fold degenerate. Obviously, the dispersion results in two band crossing points (0, 0, $\pm k^{c}_{z}$) along $\Gamma$-Z line with $k^{c}_{z}=\sqrt{\frac{M_0}{M_1}}$.

By fitting the bands of the effective model with those of the DFT calculation around $\Gamma$ point, the parameters in the effective model, for KMgBi, are given by $C_0$=-0.0794 eV, $C_1$=37.7590 eV \AA$^2$, $C_2$=16.7234 eV \AA$^2$, $M_0$=-0.0797 eV, $M_1$=-39.7258 eV \AA$^2$, $M_2$=-31.6298 eV \AA$^2$, and $a$=4.0074 eV \AA. For the sake of $|C_1|< |M_1|$, KMgBi is a
type-\uppercase\expandafter{\romannumeral1} Dirac semimetal\cite{type2}. What's more, considering the fact $\frac{|C_1|}{|M_1|}=0.9505$, KMgBi is quite near the phase transition point $\frac{|C_1|}{|M_1|}=1.0$ between the two Dirac semimetal phases. Therefore, KMgBi is located at the boundary between type-\uppercase\expandafter{\romannumeral1} and type-\uppercase\expandafter{\romannumeral2} DS phases.

Similar to KMgBi, We also extract the $k\cdot{p}$ model for RbMgBi around the $\Gamma$ point. The fitting parameters are obtained as follows: $C_0$=-0.1534 eV, $C_1$=37.9605 eV \AA$^2$, $C_2$=21.6337 eV \AA$^2$, $M_0$=-0.1122 eV, $M_1$=-36.9687 eV \AA$^2$, $M_2$=-30.0223 eV \AA$^2$ and $a$=1.1008 eV \AA. As $|C_1|> |M_1|$, the model describes a type-\uppercase\expandafter{\romannumeral2} DS \cite{type2} (see Supplementary materials). As the type-\uppercase\expandafter{\romannumeral2} DS is characterized by intriguing Fermi surfaces, we also obtain the Fermi surfaces of RbMgBi,shown in Fig.\ref{band_fs} (b). Fig.\ref{band_fs} (b) shows the Fermi surfaces at the energy of Dirac point -0.0375 eV. From Fig.\ref{band_fs} (b), we can find that the Fermi surfaces of RbMgBi consist of an electron pocket and a hole pocket, which connect each other just at the two Dirac points in the bulk Brillouin zone. This is the signature of the type-\uppercase\expandafter{\romannumeral2} DS. For the partially doped material K$_{0.5}$Rb$_{0.5}$MgBi, we find that it has already turned out to be a type-\uppercase\expandafter{\romannumeral2} DS. In this material, the $\Gamma_7$ band at $\Gamma$ point is about $35.3$meV lower than that at Z point. Here, we also want to emphasize that: for $A$MgBi ($A$=K, Rb, Cs), no matter for the type-\uppercase\expandafter{\romannumeral1} or the type-\uppercase\expandafter{\romannumeral2} DS phase, there is no other Fermi surface but for that contributed by the Dirac points; the Dirac points here are rather near the Fermi energy. Therefore, we expect that the Dirac cone and the surface states can be well measured in ARPES experiments.

\begin{figure}
\centerline{\includegraphics[width=0.45\textwidth]{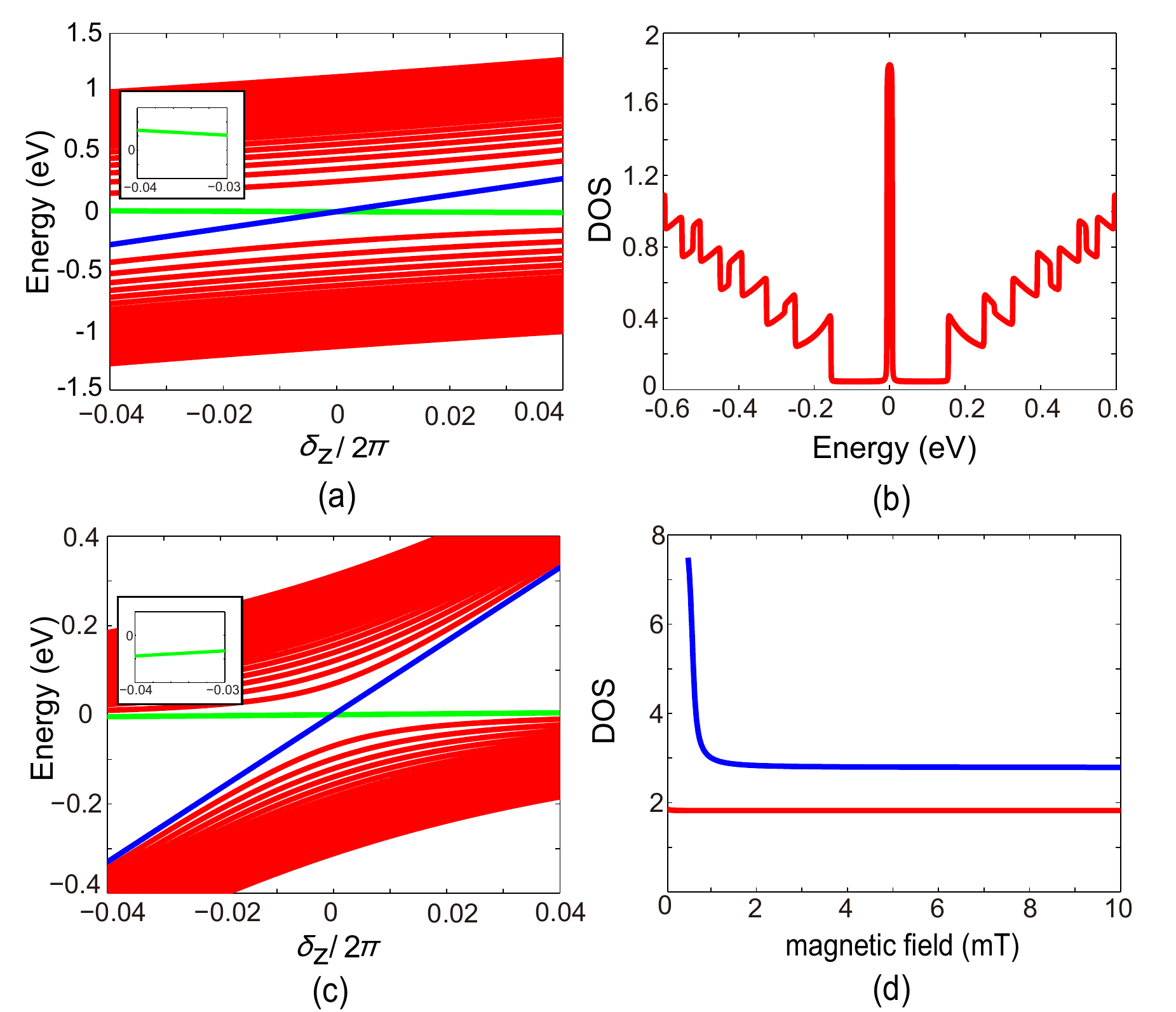}}
\caption{(color online) (a) and (c) are the Landau levels dispersion of KMgBi and RbMgBi with a magnetic field B=2 mT along the z axis, respectively. (b) The density of states of KMgBi(The density of states of RbMgBi similar to that of KMgBi).  (d)  The relation between density of states at the fermi level and magnetic field, the red (blue) line stand for KMgBi (RbMgBi).
\label{landau} }
\end{figure}

{\it Landau levels}
In the presence of magnetic field, TSs have many exotic transport properties\cite{Cr3As2 MR,BiSb MR,Narayanan2015,Feng2015,He2015,Novak2015}, which are usually closely related to the Landau levels contributed by the Weyl/Dirac points. Hence we investigate the Landau levels for the topological critical DS $A$MgBi ($A$=K,Rb, Cs). We first consider a magnetic field applied in the $xy$-plane. In earlier studies\cite{Yao2016,Udagawa,Tchoumakov,Gu2011,Lukose2007}, it was shown that the Landau levels in type-II Dirac/Weyl semimetals collapse due to open quasi-classical orbitals for field in the $xy$-plane, while in type-I Dirac/Weyl semimetals the Landau levels are well-defined. Therefore, we conclude that the Landau levels of KMgBi are well-defined but close to collapse, and in RbMgBi they will collapse due to the type-II nature of the Dirac point. When the magnetic field is applied along $z$-axis, however, the Landau levels always exist for both cases.Fig.\ref{landau} (a) and (c) show the Landau levels contributed by the Dirac point at $(0,0,k_z^c)$ for KMgBi and RbMgBi, respectively. Obviously, each Dirac point will contribute to two chiral modes corresponding to the two degenerate Weyl points. For both cases, there is one nearly nondispersive chiral mode, which is a reflection of the criticality of the bulk Dirac point. The flat chiral mode leads to a large DOS near the Fermi level, shown in Fig.\ref{landau}(b), which can be detected by STM measurement. However, there are key differences between the two cases. The two chiral modes of KMgBi have different slopes---one has a positive slope and the other a negative slope, while for RbMgBi, the slope of both chiral modes is positive, shown in Fig.\ref{landau}(a) and (c). This leads to the result that at large $k_z$, the $n\neq 0$ Landau levels can always cross the Fermi level for RbMgBi, which can not happen for KMgBi. Hence, RbMgBi has a much larger DOS at the Fermi level than KMgBi. For the fact that the $n=0$ Landau levels are independent of the magnetic field, the DOS at the Fermi level of KMgBi will not vary with the magnetic field. Nevertheless, since the energy gap between different Landau levels will increase with the increasing of magnetic field, the DOS at the Fermi level of RbMgBi will decrease rapidly, shown in Fig.\ref{landau}(d). These results can also be clearly verified by the STM measurement and the difference between KMgBi and RbMgBi can also be verified by the magneto-optical measurements\cite{Yao2016}.

{\it Discussion }
Above results suggest that  AMgBi  can be a very special family of materials to explore semimetal physics that are absent in the previously known DSs Na$_3$Bi and Cd$_3$As$_2$. Because of their special location near the boundary  of type-\uppercase\expandafter{\romannumeral1} and type-\uppercase\expandafter{\romannumeral2} Dirac semimetal phases, which is characterized by flat bands, one interesting physics is the effect of electron-electron correlation in these materials. The effect  can be strongly enhanced due to the effective dimension reduction in the electronic structure near  the boundary, which may result in intriguing many-body  emergent physics. The other is the possibility of the development of superconductivity in these materials by carrier doping. As  the materials have  identical lattice structure of the 111 iron-based superconductors, it is very likely that the carrier doping can induce superconductivity. The specific studies on these issues will be reported in the future.

In summary,  we  predict that the AMgBi are symmetry protected Dirac semimetals located near the boundary of type-\uppercase\expandafter{\romannumeral1} and type-\uppercase\expandafter{\romannumeral2} Dirac semimetal phases. The transition between the two types can be driven  by doping Rb or Cs into KMgBi. One of the features of the topological critical Dirac semimetal is that there is always a nondispersive chiral mode. These results can be well verified by STM and ARPES measurement for its clean Fermi surfaces.

The work is supported by "973" program (Grant No.
 2010CB922904 and No. 2012CV821400), as well as  national science foundation of China (Grant No. NSFC-1190024, 11175248 and 11104339).

{\it Note added}: During the revision of the present work, we notice that an experimental work on the physical properties of KMgBi \cite{Zhang2016} has been carried out.

\end{document}